\documentclass[12pt]{iopart}

\usepackage{iopams}  
\usepackage[english]{babel}
\usepackage{graphicx}
\usepackage{bbm}

\begin{document}

%
\title{The exact solution of generalized Dicke models via Susskind--Glogower operators}

\author{B.~M. Rodr\'{\i}guez-Lara and H.~M. Moya-Cessa}
\address{Instituto Nacional de Astrof\'{\i}sica, \'Optica y Electr\'onica, Calle Luis Enrique Erro No. 1, Sta. Ma. Tonantzintla, Pue. CP 72840, M\'exico}
\ead{bmlara@inaoep.mx}

%
\begin{abstract}
We show a right unitary transformation approach based on Susskind--Glogower operators that diagonalizes a generalized Dicke Hamiltonian in the field basis and delivers a tridiagonal Hamiltonian in the Dicke basis. 
This tridiagonal Hamiltonian is diagonalized by a set of orthogonal polynomials satisfying a three-term recurrence relation.
Our result is used to deliver a closed form, analytic time evolution for the case of a Jaynes--Cummings--Kerr model and to study the time evolution of the population inversion, reduced field entropy, and Husimi's Q-function of the field for ensembles of interacting two-level systems under a Dicke--Kerr model. 

\end{abstract}

\pacs{02.30.Ik,03.65.Fd,42.50.Pq}
\submitto{\JPA}

\maketitle

%
\section{Introduction}

The Jaynes--Cummings (JC) model is a fundamental building block in quantum optics; 
it describes the interaction of a qubit with a quantum electromagnetic field under long wave and rotating wave approximations. 
It is exactly solvable \cite{Jaynes1963p89} and has proven useful to describe phenomena as Rabi oscillations \cite{Allen1987} and collapse and revivals of the atomic inversion \cite{Eberly1980p1323}, among others; see~\cite{Shore1993p1195} for a review on the model.
If the number of qubits increases, the model, known as the Dicke or Tavis--Cummings model, shows many-body phenomena in the form of a superradiant phase \cite{Dicke1954p99}. 
The Dicke model is also exactly solvable \cite{Tavis1968p170, Mallory1969p1976, Hepp1973p360} and is known  to show super-fluorescence and amplified spontaneous emission; see~\cite{Garraway2011p1137} for a recent review.

In recent years, a general Dicke Hamiltonian, including quadratic self-interactions on both the field and qubit ensemble was introduced to study the effect of the nonlinearities and their relation to the Stark shift, in units of $\hbar$,
\begin{eqnarray}
H = \omega \hat{a}^{\dagger} \hat{a} + \omega_{0} \hat{S}_{z} + \gamma \left(  \hat{a}^{\dagger 2} \hat{a}^{2} + \hat{S}_{z}^2 \right)+ g \left( \hat{a}\hat{S}_{+} + \hat{a}^{\dagger} \hat{S}_{-} \right). \label{eq:HBogoliubov} 
\end{eqnarray}
In this model the frequencies for the field and two-level system transitions are given by $\omega$ and $\omega_{0}$, the quadratic interactions are assumed to be equal and given by $\gamma$, while the coupling between field and qubit is given by the parameter $g$.
An exact solution to this system was found by quantum inverse methods involving Bethe anzats~\cite{Bogoliubov1996p6305}. 
The importance of the Dicke model and its generalizations lies in its ability to describe more than atoms interacting with the quantized field of a cavity; i.e. lasers. 
For example, it may describe open dynamical cavity-QED systems~\cite{Dimer2007p013804}, ion trap systems~\cite{onen2009p033841}, circuit-QED systems~\cite{Blais2004p062320,Fink2009p083601}, and Bose-Einstein condensates interacting with classical or quantized electromagnetic fields~\cite{Chen2007p40004,Baumann2010p1301,Nagy2010p130401}.

In this contribution, we present an exact solution, up to the roots of a polynomial, to a more general Dicke Hamiltonian by considering non-identical nonlinear interactions in (\ref{eq:HBogoliubov}).
In the following, we will discuss our general Dicke Hamiltonian and the physical systems it can describe.
We then show how a novel right unitary transform involving Susskind--Glogower operators helps us diagonalize the Hamiltonian in the field basis. 
With this at hand, it is simple to diagonalize the resulting tridiagonal Hamiltonian in the Dicke basis via orthogonal polynomials satisfying a three-term recurrence relation.
In order to verify the validity of our exact solution, we recover the time evolution for a system involving just the single qubit. 
Finally, we study the time evolution of different ensemble sizes to illustrate the simplicity of our approach and the results it yields; we focus on the population inversion dynamics of the qubit ensemble as well as the evolution of the entropy and Q-function of the field.

%
\section{The model}
Let us consider a system composed by an ensemble of $N$ identical two-level systems (`qubits') that interact with each other. These qubitas are in the presence of a quantized field and a Kerr medium. For the sake of simplicity, we move into the frame defined by the transformation $\hat{U}(t) = e^{- i \omega_{f} \hat{N} t}$, where the excitation number operator is given by $\hat{N}=\hat{a}^{\dagger} \hat{a} + \hat{J}_{z}$ , and work with the Hamiltonian in units of $\hbar$,
\begin{eqnarray}  \label{eq:Hamiltonian}
\hat{H} =\delta \hat{J}_{z} + \kappa \left( \hat{a}^{\dagger} \hat{a} \right)^2 +  \gamma \hat{J}_{z}^{2}+ \lambda \left( \hat{a} \hat{J}_{+} + \hat{a}^{\dagger} \hat{J}_{-} \right).
\end{eqnarray}
The qubits ensemble is described by collective Dicke operators satisfying the $su(2)$ algebra, $\left[ \hat{J}_{+}, \hat{J}_{-} \right] = 2 \hat{J}_{z}$, $\left[ \hat{J}_{z}, \hat{J}_{\pm} \right] = \pm \hat{J}_{\pm}$, while the annihilation and creation operators for a single mode field satisfy $\left[ \hat{a}, \hat{a}^{\dagger}\right]=1$.
The transition frequency of each qubit, $\omega_{q}$, and the frequency of the field, $\omega_{f}$, are summarized by the detuning  $\delta = \omega_{q} - \omega_{f}$.
The Kerr medium is described by the parameter $\kappa$, while the qubit-qubit and ensemble-field couplings are given by $\gamma$ and $\lambda$, in that order.

The Hamiltonian (\ref{eq:Hamiltonian}) describes the $N$-atom maser in general. 
In the special case of equal self-interactions, $\kappa = \gamma$, it can be transformed into the $N$-atom maser including, Kerr nonlinearity and Stark shift as discussed in~\cite{Bogoliubov1996p6305}.
Different parameter sets describe particular physical models; e.g.,$\{\delta, \gamma, \lambda\}=0$ delivers the Kerr model~\cite{Drummond1980p725, Flytzanis1980p441},  $\{ \kappa, \gamma \} = 0$ yield the Dicke or Tavis--Cummings model~\cite{Dicke1954p99, Tavis1968p170} and $\{ \gamma \} = 0$ gives the micromaser with Kerr nonlinearity~\cite{Deb1993p3191, Guo2011p1245, Guo12p023608}.
Furthermore, the general Hamiltonian (\ref{eq:Hamiltonian}) and its reductions are experimentally feasible in cavity- and circuit-QED as well as trapped ions.
It may also be possible to realize some of these models with two-mode Bose-Einstein condensates coupled to radiation fields~\cite{ Chen2007p40004, Chen2008p023634,  RodriguezLara2010p2443, RodriguezLara2011p016225, RodriguezLara2012}.

The case of equal-self interactions, $\kappa = \gamma$, has been solved by inverse quantum methods in the past \cite{Bogoliubov1996p6305}.
This solution involves the Bethe ansatz method.
The general Hamiltonian (\ref{eq:Hamiltonian}) can also be solved by extending our right unitary approach to the quantum Landau--Zener problem for a single two-level system presented in~\cite{RodriguezLara2011p3770}, which delivers an evolution operator with the form
\begin{eqnarray}
\hat{U}\left(t\right) &=& \hat{U}_{A}\left(t\right) \hat{U}_{B}\left(t\right), \qquad \hat{U}_{x} =  e^{-i \hat{H}_{x} t},
\end{eqnarray}
where the auxiliary Hamiltonians are given by
\begin{eqnarray}
\hat{H}_{A} &=& \sum_{j=-N/2}^{N/2} F\left(j,\hat{n}\right) \vert j \rangle \langle  j \vert + \sum_{j=-N/2+1}^{N/2} G\left(j,\hat{n}\right) \left( \vert j \rangle \langle j-1 \vert + \vert j-1 \rangle \langle j \vert \right) ,  \nonumber \\~\\
\hat{H}_{B} &=& \sum_{j=-N/2}^{N/2-1} \sum_{k=0}^{N/2-1-j} F\left(j,\hat{n}\right) \hat{\rho}_{k} \vert j  \rangle \langle j  \vert + \nonumber \\
&&+ (1 - \delta_{N,1})\sum_{j=-N/2+1}^{N/2-1} \sum_{k=0}^{N/2-1-j} G\left(j,\hat{n}\right) \hat{\rho}_{k} \left( \vert j \rangle \langle j-1 \vert + \vert j-1 \rangle \langle j \vert \right), \nonumber \\~
\end{eqnarray}
where the ket $\vert j \rangle$ is a Dicke state, the operator $\hat{\rho}_{k}$ is the density matrix for the pure state of the field with $k$ photons, the operator $\hat{n} = \hat{a}^{\dagger} \hat{a}$ is the photon number operator and the symbol $\delta_{a,b}$ represents Kronecker delta.
These auxiliary Hamiltonians are diagonal in the field basis; i.e. they are given in terms of the photon number functions 
\begin{eqnarray}
F\left(j,\hat{n}\right) &=& \kappa   \left( \hat{n} - \frac{N}{2} + j \right)^2  + j \left( \delta  + \gamma j \right),  \\
G\left(j,\hat{n}\right) &=& \lambda \left[ \frac{N}{2} \left(\frac{N}{2} + 1 \right) - j \left( j - 1 \right) \right]^{1/2} \left[ \hat{n} + 1 + \frac{N}{2} - j \right]^{1/2}.
\end{eqnarray}
There is, however, a simpler approach to solve this general radiation-matter interaction model.

%
\section{Exact solution}

In order to present a simpler approach to solve Hamiltonian (\ref{eq:Hamiltonian}), let us define the right unitary transformation
\begin{eqnarray}
\hat{T} &=& \sum_{j=-\frac{N}{2}}^{\frac{N}{2}} \hat{V}^{\frac{N}{2} + j} \vert j \rangle \langle j \vert,\\
\hat{T} \hat{T}^{\dagger} &=& \mathbbm{1} , \\
\hat{T}^{\dagger} \hat{T} &=& \mathbbm{1} -  \sum_{j=-\frac{N}{2}+1}^{\frac{N}{2}} \sum_{k=0}^{\frac{N}{2}-1+j} \hat{\rho}_{k} ~\vert j \rangle \langle j \vert, \quad \hat{\rho}_{k} = \vert k \rangle_{ff}\langle k \vert,  \label{eq:RightUnitary}
\end{eqnarray}
where we have used the Susskind--Glogower operators,
\begin{eqnarray}
\hat{V} = \frac{1}{\sqrt{\hat{a}^{\dagger} \hat{a} +1}} ~\hat{a}, \\
\hat{V}^{\dagger} = \hat{a}^{\dagger} ~\frac{1}{\sqrt{\hat{a}^{\dagger} \hat{a} +1}} , 
\end{eqnarray}
which act as lowering and raising ladder operators on the Fock state basis, $\hat{V} \vert n \rangle_{f} = \vert n-1 \rangle_{f} $ and $\hat{V}^{\dagger} \vert n \rangle_{f} = \vert n+1 \rangle_{f} $ in that order, and are right-unitary, $\hat{V}\hat{V}^{\dagger} = 1$ and $\hat{V}^{\dagger} \hat{V} = 1 - \hat{\rho}_{0}$, where $\hat{\rho}_{k}$ is the density matrix for the pure state of the field with $k$ photons. Again, the ket $\vert j \rangle$ is a Dicke or angular momentum state.
Then, it is possible to write the general Hamiltonian (\ref{eq:Hamiltonian}) as:
\begin{eqnarray}
\hat{H} = \hat{T} \hat{H}_{SC} \hat{T}^{\dagger},
\end{eqnarray}
where the auxiliary Hamiltonian is given by, 
\begin{eqnarray}
\hat{H}_{SC} &=& \sum_{j=-\frac{N}{2}}^{\frac{N}{2}} f\left(j,\hat{n}\right) \vert j \rangle \langle  j \vert +  \nonumber \\
&&\sum_{j=-\frac{N}{2}+1}^{\frac{N}{2}} g\left(j,\hat{n}\right) \left( \vert j \rangle \langle j-1 \vert + \vert j-1 \rangle \langle j \vert \right) . \label{eq:SCHam}
\end{eqnarray}
We have used the notation $\hat{H}_{SC}$ to bring forward that this Hamiltonian is \textit{semi-classical}-like because it is only expressed in terms of the number operator,
\begin{eqnarray}
f\left(j,\hat{n}\right) &=& \kappa   \left( \hat{n} - \frac{N}{2} - j \right)^2  + j \left( \delta  + \gamma j \right),  \\
g\left(j,\hat{n}\right) &=& \lambda \left[ \frac{N}{2} \left(\frac{N}{2} + 1 \right) - j \left( j - 1 \right) \right]^{1/2} \left[ \hat{n} + 1 - \frac{N}{2} -j \right]^{1/2}. 
\end{eqnarray}
It is possible to express the dynamics of this model as the evolution operator
\begin{eqnarray}
\hat{U}\left(t\right) = e^{- \imath t \hat{H}} = \sum_{m} \frac{1}{m!} \left(-\imath t \hat{H}\right)^m ,
\end{eqnarray}
where powers of the form $\hat{H}^{m}$ are needed. 
These powers can be obtained by realizing from (\ref{eq:RightUnitary}) and (\ref{eq:SCHam}) that $  \hat{T}  \hat{H}_{SC} \hat{T}^{\dagger} \hat{T} \hat{H}_{SC}  \hat{T}^{\dagger} =  \hat{T} \hat{H}_{SC}^2  \hat{T}^{\dagger}$ leads to $\hat{H}^m = \hat{T} \tilde{H}^m \hat{T}^{\dagger}$ by means of $ \hat{V}^{\frac{N}{2} + j} \vert k \rangle_{f} = 0 $ and $  _f \langle k \vert (\hat{V}^{\dagger})^{\frac{N}{2} + j}= 0$ for $k=0,\ldots,N/2 +j -1 $ and $j = -N/2 + 1, \ldots, N/2$.
Thus, the evolution operator in the reduced form is given by the expression
\begin{eqnarray}
\hat{U}\left(t\right) = \hat{T} e^{-\imath t \hat{H}_{SC} } \hat{T}^{\dagger}.
\end{eqnarray}
The Hamiltonian $\hat{H}_{SC}$ is diagonal in the field basis and is symmetric tridiagonal in the Dicke basis; i.e. it is diagonalizable in the angular momentum basis. 
The eigenvalues of this Hamiltonian can be found by the method of minors and are given by the roots of the characteristic polynomial 
\begin{eqnarray}
p_{N+1}\left(\nu\right) &=& \left[ \nu - f\left(-\frac{N}{2},\hat{n}\right)  \right] p_{N}\left(\nu\right) - g^2\left(-\frac{N}{2}+1,\hat{n}\right)  p_{N-1}\left(\nu\right)
\end{eqnarray}
with
\begin{eqnarray}
p_{0}\left(\nu\right) &=& 1 ,  \\
p_{1}\left(\nu\right) &=& \nu - f\left(\frac{N}{2},\hat{n}\right)  ,  \\
p_{j}\left(\nu\right) &=& \left[ \nu - f\left(\frac{N}{2}+1 -j,\hat{n}\right)  \right] p_{j-1}\left(\nu\right) + \nonumber \\
&&  - g^2\left(\frac{N}{2} + 2 -j, \hat{n}\right)  p_{j-2}\left(\nu\right), \quad j \ge 2 \nonumber \\
\end{eqnarray}
The eigenvectors are calculated from the eigenvalue equation for the Hamiltonian and give
\begin{eqnarray}
\vert v_{j} \rangle &=& \sum_{k=-\frac{N}{2}}^{\frac{N}{2}} c_{k}^{\left(j\right)} \vert k \rangle, \qquad \sum_{k=-\frac{N}{2}}^{\frac{N}{2}} \vert c_{k}^{\left(j\right)}\vert^2 =1,
\end{eqnarray}
where the amplitudes answer to the following recurrence relations,
\begin{eqnarray}
\left[ f\left( \frac{N}{2}, \hat{n} \right) - \nu_{j} \right] c_{\frac{N}{2}}^{\left(j\right)}  +  g\left( \frac{N}{2}, \hat{n} \right) c_{\frac{N}{2} - 1}^{\left(j\right)}&=& 0 ,  \\
\left[ f\left(j, \hat{n} \right) - \nu_{j} \right] c_{k}^{\left(j\right)}  +  g \left(j, \hat{n} \right) c_{k- 1}^{\left(j\right)} +  g \left(j+1, \hat{n} \right) c_{k + 1}^{\left(j\right)} &=& 0,  \\
\left[ f\left(-\frac{N}{2}, \hat{n} \right) - \nu_{j} \right] c_{-\frac{N}{2}}^{\left(j\right)}  +  g\left(-\frac{N}{2}+1, \hat{n} \right) c_{\frac{N}{2} + 1}^{\left(j\right)}&=& 0. 
\end{eqnarray}

%
\section{Examples}

The time evolution given in the previous section accounts for the full dynamics of the system and helps calculating any given quantity of interest. 
As an example, we will focus on the time evolution of the reduced density matrix for the field where the initial state is given by a pure state $\vert \psi\left(0\right) \rangle = \sum_{j=0}^{\infty} \sum_{k=-\frac{N}{2}}^{\frac{N}{2}}  c_{j,k} \vert j \rangle_{f} \vert k \rangle$,
\begin{eqnarray}
\rho_{f}\left(t\right) &=& \sum_{j,k,l=-\frac{N}{2}}^{\frac{N}{2}} \sum_{p,q=0}^{\infty}  c_{p+l-j,j} c_{q+l-k,k}^{\ast} U_{l,j}\left(p+l+\frac{N}{2},t\right) \times \nonumber \\
&& \times  U_{l,k}^{\ast}\left(q+l+\frac{N}{2},t\right)   \vert p \rangle_{ff}\langle q \vert .
\end{eqnarray}
The notation $U_{i,j}\left(\hat{n},t\right) =  \left( e^{-\imath t \hat{H}_{SC} } \right)_{i,j}$ is used to describe the components of the \textit{semi-classical} time evolution operator.
This allows us to calculate the mean photon number evolution, 
\begin{eqnarray}
\langle \hat{n}\left(t\right) \rangle &=& \sum_{j,k,l=-\frac{N}{2}}^{\frac{N}{2}} \sum_{p=0}^{\infty} ~p~ c_{p+l-j,j} c_{p+l-k,k}^{\ast} U_{l,j}\left(p+l+\frac{N}{2},t\right) U_{l,k}^{\ast}\left(p+l+\frac{N}{2},t\right) , \nonumber \\
\end{eqnarray}
and in consequence the population inversion $\langle \hat{J}_{z}(t) \rangle = \langle \hat{N}(t=0) \rangle - \langle \hat{n}\left(t\right) \rangle$. 
Other interesting quantities are the purity of the field, 
\begin{eqnarray}
P\left(t\right) &=& 1 - \mathrm{Tr}~ \hat{\rho}_{f}^2 , \\
\mathrm{Tr}~ \hat{\rho}^2&=& \sum_{j,k,l,m,n,o=-\frac{N}{2}}^{\frac{N}{2}} \sum_{p,q=0}^{\infty}  c_{p+l-j,j} c_{q+o-m,m} c_{q+l-k,k}^{\ast} c_{p+o-n,n}^{\ast} ~U_{o,j}\left(p+l+\frac{N}{2},t\right)\times \nonumber \\
&&  U_{o,m}\left(q+o+\frac{N}{2},t\right) U_{l,k}^{\ast}\left(q+l+\frac{N}{2},t\right) U_{o,n}^{\ast}\left(p+o+\frac{N}{2},t\right), 
\end{eqnarray}
and von Neumann entropy, 
\begin{eqnarray}
\langle \hat{S}_{f}\left(t\right) \rangle = - \mathrm{Tr} \left[ \hat{\rho}_{f}\left(t\right) \ln \hat{\rho}_{f}\left(t\right) \right], 
\end{eqnarray}
which are a good measure of the degree of mixedness of the reduced system.

\subsection{A Single qubit}

\begin{figure}
\center\includegraphics[scale=1]{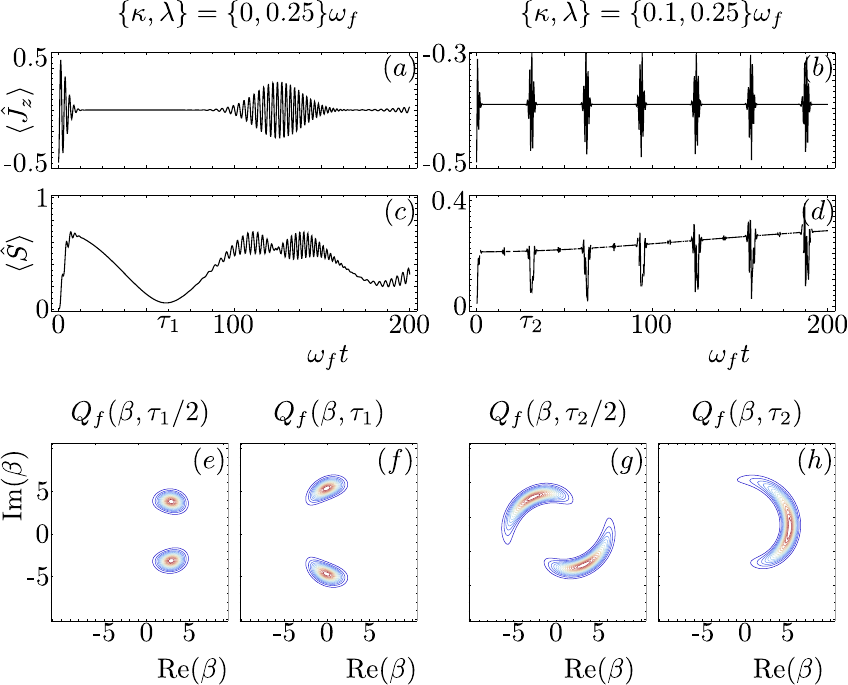}
\caption{Time evolution of the mean population inversion (a,b), reduced field entropy (c,d) and Husimi's Q-function for the field at times equal to half-minimum (e,g) and minimum of entropy (f,h) for a single qubit interacting with a quantized electromagnetic field on resonance, $\delta=0$, under the Jaynes-Cummings model, left column (a,c,e,f), and under a Jaynes-Cummings-Kerr model, right column (b,d,g,h). The initial state for both cases is $\vert \psi(0) \rangle = \vert \alpha \rangle_{f} \vert -\frac{1}{2} \rangle$ with $\alpha = 5$. }
\label{fig:Fig1}
\end{figure}

Let us consider a system with just the single qubit,
\begin{eqnarray}
\hat{H} &=& \kappa \hat{n}^2 + \frac{\delta}{2} \hat{\sigma}_{z} + \lambda \left( \hat{a} \hat{\sigma}_{+} + \hat{a}^{\dagger} \hat{\sigma}_{-}  \right) , 
\end{eqnarray}
the \textit{semi-classical} Hamiltonian is given by
\begin{eqnarray}
\hat{H} &=& \left( \begin{array}{cc} \kappa\left( n-1\right)^2 + \frac{\delta}{2} & \lambda \sqrt{\hat{n}} \\ \lambda \sqrt{\hat{n}} & \kappa  n^2 - \frac{\delta}{2},
\end{array} \right)
\end{eqnarray}
and it is possible to give a closed form time evolution operator as
\begin{eqnarray}
\hat{U}(t) &=& \hat{T} e^{- i t \hat{H}_{SC}} \hat{T}^{\dagger}, \\
e^{- i t \hat{H}_{SC}} &=& e^{- \frac{\imath t}{2} \kappa \left[ 1 + 2 \hat{n} \left( \hat{n} + 1 \right)\right]} \left\{ \cos \frac{\Omega(\hat{n}) t}{2} - \frac{i \left[\beta(\hat{n}) \hat{\sigma}_{z} + 2 \lambda \sqrt{n} \hat{\sigma}_{x} \right]}{\Omega(\hat{n})} \sin \frac{\Omega(\hat{n})t}{2} \right\},\\
\beta(\hat{n}) &=& \delta + \kappa \left(1 - 2 \hat{n} \right), \\
\Omega(\hat{n}) &=& \sqrt{ \left[ \beta(\hat{n}) \right]^2 + 4 \hat{n} \lambda^2 }
\end{eqnarray}
It is trivial to apply the operator $\hat{T}^{\dagger}$ ($\hat{T}$) to any given initial state ket (bra) and then apply the \textit{semi-classical} exponential. 
Figure \ref{fig:Fig1} shows the time evolution of the mean population inversion (first row), entropy of the reduced field (second row) and Husimi's Q--function of the field (third row) for a single qubit as given by a Jaynes--Cummings model (left column) and a Jaynes--Cummings--Kerr model (right column).
Our results are in accordance with those in the literature \cite{Eberly1980p1323, MoyaCessa1995p51} and we can proceed to sample the dynamics of ensembles.

\subsection{An ensemble of qubits}

\begin{figure}
\center\includegraphics[scale=1]{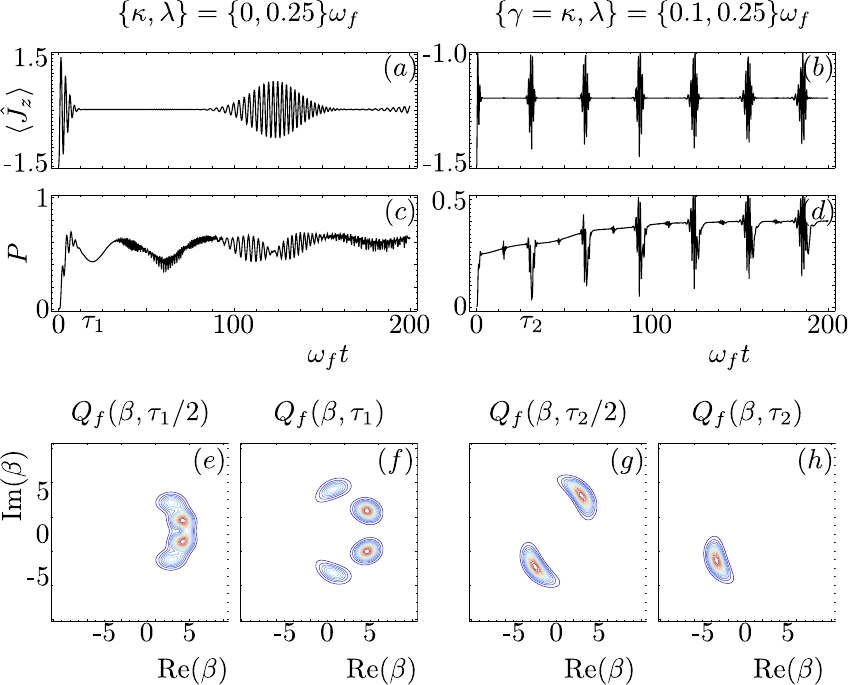}
\caption{ Time evolution of the mean population inversion (a,b), reduced field purity (c,d) and Husimi's Q-function for the field at times equal to half-minimum (e,g) and minimum of entropy (f,h) for a quantized electromagnetic field interacting on resonance, $\delta=0$, with three qubits under the Dicke model, left column (a,c,e,f),  and with three interacting qubits under a Dicke-Kerr model, right column (b,d,g,h). The initial state for both cases is $\vert \psi(0) \rangle = \vert \alpha \rangle_{f} \vert -\frac{3}{2} \rangle$ with $\alpha = 5$.}
\label{fig:Fig2}
\end{figure}

\begin{figure}
\center\includegraphics[scale=1]{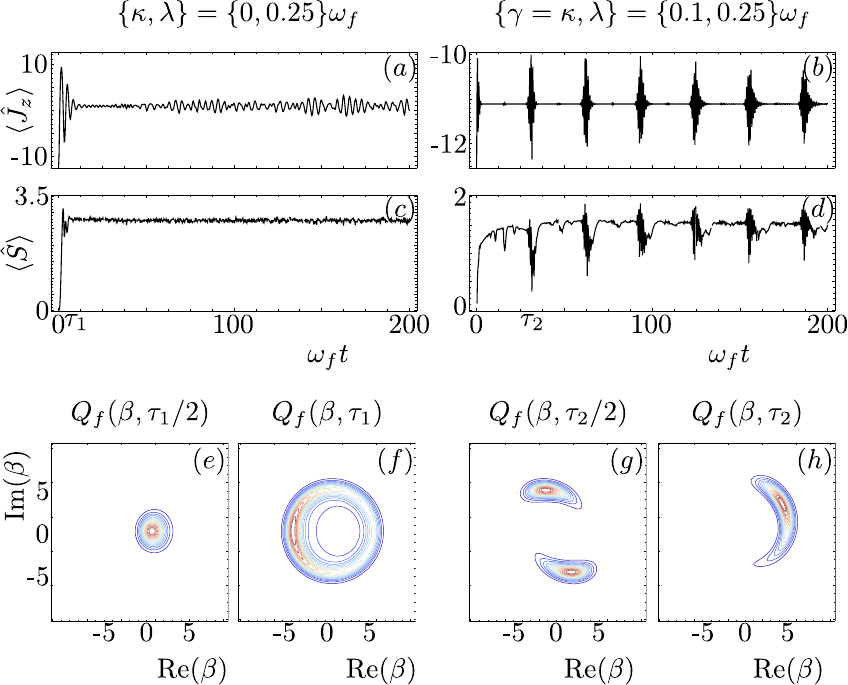}
\caption{ Time evolution of the mean population inversion (a,b), reduced field entropy (c,d) and Husimi's Q-function for the field at times equal to half-minimum (e,g) and minimum of entropy (f,h) for a quantized electromagnetic field interacting on resonance, $\delta=0$, with twenty five qubits under the Dicke model, left column (a,c,e,f), and with twenty five interacting qubits under a Dicke-Kerr model, right column (b,d,g,h). The initial state for both cases is $\vert \psi(0) \rangle = \vert \alpha \rangle_{f} \vert -\frac{25}{2} \rangle$ with $\alpha = 5$.}
\label{fig:Fig3}
\end{figure}

For an ensemble of qubits, the task of finding a closed form expression for the time evolution becomes cumbersome but it is possible to numerically diagonalize the \textit{semi-classical} Hamiltonian and implement the time evolution of any given initial state.
As an example, we consider the evolution of ensembles of three, Fig. \ref{fig:Fig2}, and twenty five, Fig. \ref{fig:Fig3}, qubits. 
The information about the particular initial conditions and parameter values can be found in the figures and their captions.
At the time, it is not our goal to report and in-depth analysis of the dynamics of generalized Dicke models but just to present our diagonalization scheme to obtain an exact solution via Susskind-Glogower operators. 
For this reason, we will just briefly comment some basic characteristics of the dynamics.
By considering an initial state given by the separable state consisting of a coherent field and the ensemble in its ground state,  $\vert \psi(0) \rangle = \vert \alpha \rangle_{f} \vert -N/2 \rangle$, it is possible to see that the Dicke model presents strong collapse and revivals of the population inversion as long as the mean photon number is larger than the number of qubits in the system. 
A clear collapse of the population inversion is seen in any case studied here, up to $N \sim \alpha^2$.
The strength of the oscillations in the population inversion diminishes as the number of qubits in the system gets close to the mean photon number of the coherent state but they become ever-present at smaller times as we get larger ensemble sizes for a fixed value of the coherent state parameter.
Meanwhile, the purity and entropy of such a Dicke model signals an ever-present entangled state between the field and the ensemble as the number of qubits gets close to or equal to the mean number of photons; i.e. the plots change from strong, well-defined, unmodulated dips in the functions to a strongly modulated flat-liner close to the value of a mixed reduced density matrix~\cite{Phoenix1991p6023, Zyczkowski1998p883, Zyczkowski1999p3496}.
The Q-function for the reduced field behaves as expected. 
For $\alpha \ll N$,  $N+1$ well-defined phase blobs appear and evolve half of them clock-wise and the other half counter-clock-wise as time goes by. 
The revivals in the population inversion are associated to the overlapping of these phase blobs; a stronger revival corresponding to a better overlapping. 

However, when an interacting ensemble of qubits is considered under Dicke--Kerr dynamics, the collapse and revivals of the population inversion are always weak but well defined and periodical. 
Purity and entropy functions point a return to a quasi-separable state on the first revival for the cases analyzed with the number of qubits less or equal to the mean photon number of the field.
The mean value of these functions gradually increases with time and some dips appear periodically due to the constructive interference of the wavefunction components, leading to revivals in the population inversion.
Under Dicke--Kerr dynamics the phase blobs seem heavily defined by the Kerr process and for $\alpha = 5$ four phase blobs appear and two of them evolve clockwise while the other two do it counter-clockwise. 
This process produces an overlap of two and two of the phase blobs leading to a weak local minimum in the purity/entropy but does not register in the population inversion.
It is only when the four phase blobs overlap that a pronounced local minimum and a revival of the population inversion appears.

%
\section{Conclusion}

We have considered the general $N$-atom maser model which can be described by the Dicke model plus dipople--dipole interactions and Kerr nonlinearity. 
As a side result, we extend a previous result based on Susskind--Glogower operators that gives the exact dynamics of a Jaynes--Cummings model as the product of two time evolution operators.
Our main result is a different and simpler approach involving Susskind--Glogower operators and right unitary transformations that allow us to represent our generalized Dicke model as a transformed \textit{semi-classical}-like Hamiltonian which is diagonal in the field basis and tridiagonal in the Dicke basis; thus, the diagonalization of this \textit{semi-classical} Hamiltonian is known up to the roots of its characteristic polynomial. 
The transformed \textit{semi-classical}-like Hamiltonian gives the time evolution of the system and provides access to the dynamics of any quantity of interest.

We use our result to derive a closed analytical form for the time evolution operator of a single qubit interacting with a quantized field in the presence of a Kerr medium, a Jaynes-Cummings-Kerr model. 
Also, we present the time evolution of the population inversion, reduced field entropy and Husimi's Q-function of the field for ensembles consisting of three and twenty-five interacting two-level systems under a Dicke-Kerr model where the interaction and Kerr nonlinearity are equal.
This is done to show how simple it is to deal with many atoms with our partial diagonalization approach.

It is possible that one could follow the dynamics of hundreds and maybe a few thousands of qubits  with our approach in a simple workstation with efficient programming; e.g., this is of importance in the description of realistic micromasers and may be relevant to the study of fields interacting with Bose-Einstein condensates in the two-mode approximation.

%

\appendix
\section{Small rotations for a generalized quantum Rabi model}

Some systems, e.g. circuit-QED and open-dynamical systems, may deliver a strong coupled version of the general Dicke Hamiltonian in (\ref{eq:Hamiltonian}), 
\begin{eqnarray} \label{eq:FullHamiltonian}
H &=& \hat{H}_{0} + \hat{H}_{I}, \nonumber \\
\hat{H}_{0} &=& \omega_{f} \hat{a}^{\dagger} \hat{a} + \kappa \left( \hat{a}^{\dagger} \hat{a} \right)^2 + \chi \left(\hat{a}^{2} + \hat{a}^{\dagger 2}\right) + \omega_{q} \hat{J}_{z} + \frac{\xi}{N} \hat{J}_{z}^{2} ,\nonumber \nonumber \\
\hat{H}_{I}&=&\frac{g}{\sqrt{N}} \left( \hat{a} + \hat{a}^{\dagger} \right) \left(\hat{J}_{+} + \hat{J}_{-} \right).
\end{eqnarray}
Notice that the $\mathbf{A}^2 \propto \left( \hat{a} + \hat{a}^{\dagger} \right)^2$ term \cite{Dicke1954p99} has been kept for the sake of generality. 
The presence of the strong interaction term deters the use of the approach presented above. 

Here, we want to show two things. 
The first is that we can get rid of the second order nonlinearity, $\chi$, if it is weak compared to the frequency of the field. 
This allows us to use a squeezed states basis for the field, described by the transformation,
\begin{eqnarray}
\hat{T}_{1} &=& e^{\frac{\chi}{2 \omega_{f}}  \left(\hat{a}^{2} - \hat{a}^{\dagger 2}\right) }, \qquad  \frac{\chi}{\omega_{f}} \ll 1.
\end{eqnarray}
that helps us get rid of the $\chi$ term.
The second thing we want to show is that a \textit{small rotation}~\cite{Klimov1999p063802},
\begin{eqnarray}
\hat{T}_{2} =  e^{ \frac{\tilde{g}}{\tilde{\omega}_{f} + \omega_{q}} \left(\hat{a} - \hat{a}^{\dagger} \right) \left( \hat{J}_{+} + \hat{J}_{-} \right)}, \quad \frac{\tilde{g}}{\tilde{\omega}_{f} + \omega_{q}} \ll 1,
\end{eqnarray}
has an effect similar to that of the rotating wave approximation.
This \textit{small rotation} leads to just a Dicke Hamiltonian including a Kerr medium and dipole-dipole interactions between the qubit ensemble components,
\begin{eqnarray}  \label{eq:WeakHamiltonian}
\hat{H} =\delta \hat{J}_{z} + \kappa \left( \hat{a}^{\dagger} \hat{a} \right)^2 +  \gamma \hat{J}_{z}^{2}+ \lambda \left( \hat{a} \hat{J}_{+} + \hat{a}^{\dagger} \hat{J}_{-} \right),
\end{eqnarray}
after we have moved to a frame defined by the total excitation number $\hat{N} = \hat{a}^{\dagger} \hat{a} + \hat{J}_{z}$ rotating at the frequency of the field and defined the parameters $\delta = \omega_{q} - \omega_{f} + 2 \chi^2 / \omega_{f}$, $\gamma = \xi / N$ and $\lambda = 2 g \left(\omega_{f} - \chi\right) \left(\omega_{f}^2 - 2 \chi^2\right) / \sqrt{N} \omega_{f} \left(\omega_{f}^2 - 2 \chi^2 + \omega_{q} \omega_{f}\right)$.
Note that we have taken the self-interaction nonlinearities $\kappa$ and $\xi$ a couple orders of magnitude smaller than the transition frequency $\omega_{q}$ in order to neglect products of couplings and nonlinearities.

We want to emphasize that, while we cannot deal with the strong-coupling regime, this \textit{small rotation} method may be valid in the regime where phase transitions appear $g_{c} = \sqrt{ \left( \omega_{f} - 2 \chi^{2}/\omega_{f} \right) \left(\omega_{q}- \xi\right)}$ \cite{RodriguezLara2010p2443,RodriguezLara2010}.

%
\section*{References}

\providecommand{\newblock}{}

\end{document}